\documentclass{iau}

\usepackage[T1]{fontenc}
\usepackage[utf8]{inputenc}
\usepackage{amsmath}
\usepackage{graphicx}
\usepackage{multirow}
\usepackage{caption}    
\usepackage{capt-of}
\usepackage{wrapfig}

\begin{document}

\lefttitle{Debatri Chattopadhyay et al.}
\righttitle{Evolutionary Links: From Gaia Neutron Star Binaries to Pulsar White Dwarf Endpoints}

\jnlPage{1}{7}
\jnlDoiYr{2021}
\doival{10.1017/xxxxx}

\aopheadtitle{Proceedings IAU Symposium}
\editors{Editors.}

\title{Evolutionary Links: From Gaia Neutron Star Binaries to Pulsar White Dwarf Endpoints}

\author{Debatri Chattopadhyay$^{1}$, Kyle A. Rocha$^{1}$, Seth Gossage$^{1}$ and Vicky Kalogera$^{1}$}
\affiliation{$^{1}$Center for Interdisciplinary Exploration and Research in Astrophysics (CIERA) and Department of Physics \& Astronomy, Northwestern University, 1800 Sherman Ave, Evanston, IL 60201, USA}

\begin{abstract}
The discovery of wide, eccentric Gaia neutron stars (NSs) in binaries with still evolving (likely main sequence) companions offers a new probe of mass transfer and pulsar recycling beyond the compact-binary regime. We model their origins and fates using population synthesis with \texttt{POSYDON} and detailed binary evolution with \texttt{MESA}, contrasting two limiting prescriptions at Roche-lobe overflow (RLOF): enforced circularization versus explicitly eccentric mass transfer. Our \texttt{MESA} setups include updated treatments of eccentric, non-conservative transfer, magnetic-braking torques for cool stars, and neutron-star spin evolution with accretion and dipole spindown. Under optimistic assumptions, isolated evolution yields Gaia-like systems at only $\leq$1.5\% relative rates of NS-evolving companion binaries, yet absolute numbers remain consistent with detections for continuous star formation. Synthetic populations indicate that many canonical millisecond pulsar–white dwarf (WD) binaries arise from unstable mass transfer and common envelope recycling, whereas Gaia systems typically avoid common envelope and only undergo stable mass transfer. In the case of capping accretion onto the NS at the Eddington rate, circular RLOF keeps the donor’s mass-loss rate hovering around the Eddington limit and sustained over long timescales. Eccentric mass transfer instead produces briefer burst-y signature where the donor’s mass-loss rate can climb up to a thousand times higher than in the circular case. The eccentric channel then leaves wide, eccentric NS–helium WD binaries with only mild recycling, whereas the circular channel yields long-lived transfer, circular NS–WD binaries (helium or carbon-oxygen core), and fully recycled millisecond pulsars.

\end{abstract}


\maketitle

\section{Introduction}

The discovery of a new population of wide, eccentric neutron star(NS)-main-sequence(MS) binaries by Gaia has opened a novel window to binary evolution. In Gaia Data Release 3, astrometric orbital solutions revealed dozens of solar-type stars wobbling under the gravity of unseen companions of mass $M_{\rm comp}\sim1.3$–$1.8\,M_\odot$ – consistent with NS rather than white dwarfs (WDs). Follow-up spectroscopy confirmed 21 NS–MS binaries in the Galactic field \citep{ElBadry2024}. These systems have orbital periods $P_\mathrm{orb}$ of roughly 200 to 1,000 days and surprisingly high eccentricities ($e\simeq0.2$–0.8), in stark contrast to the nearly circular orbits of typical NS (as millisecond pulsar or MSP)-WD binaries of similar period, discovered through radio astronomy \citep{Manchester2005}. Moreover, a few primaries show unusual characteristics (e.g. Lithium-enhancement and low metallicity), hinting at exotic formation channels. These findings challenge the long-held assumption that Roche-lobe overflow (RLOF) in binary stars leads to prompt orbital circularization \citep{Hut1981}, an assumption that underpins most binary evolution models \citep{Hurley2002}. If mass transfer can occur while the orbit remains eccentric, the future evolution of these NS–MS binaries could differ substantially from the standard picture of NS recycling.

In this contribution, we provide an overview of key results on the future evolution of Gaia’s NS–MS binaries, emphasising the contrast between eccentric mass transfer vs usual assumption of artificial circular RLOF outcomes. We limit our scope to the essential findings, which serve as a conceptual teaser, while deferring detailed modelling aspects to forthcoming journal article(s). In particular, we focus on how each Gaia binary will likely end up as a NS-WD system (rather, a radio pulsar–WD pair), and how the nature of mass transfer influences the final orbital and stellar properties. We also connect our predictions to the observed population of Galactic pulsar–WD binaries, which comprises $\sim$130 systems in the field, to see which, if at all, evolutionary channel (eccentric vs.\ circular) is actually responsible for their formation.

\section{Methods}
To assess whether the Gaia NS–MS binaries can be produced through isolated binary evolution, we first performed a population synthesis using the open-source binary evolution code \texttt{POSYDON} \citep{AndrewsPOSYDONv22024}. We explored a broad
range of initial conditions and physical assumptions (e.g. star formation history, metallicity, NS natal kicks, common envelope prescriptions) to search for progenitor binaries that would evolve into systems matching the observed Gaia sample (i.e. a $\sim$1.3–1.8$\,M_\odot$ NS with a $\sim$1\,$M_\odot$ companion in a
wide, $P_\mathrm{orb}\sim10^3$\,d, eccentric orbit).

To predict the future evolution of each Gaia NS–MS binary, we carried
out detailed binary stellar evolution calculations using the \texttt{MESA} stellar evolution code \citep{PaxtonMESA2011}. We initialized 1-D stellar
models for each of the 21 Gaia systems. The $M_\mathrm{comp}\approx0.9$–1.8\,$M_\odot$ companion stars (using their estimated masses, orbital
periods, eccentricities, and metallicities from the Gaia observations) and evolved them together with a $1.3$–
$1.8\,M_\odot$ NS using \texttt{MESA}’s binary module. We incorporated several important physics updates: (i) an eccentric mass-transfer prescription based on \cite{Sepinsky2009} and \cite{Rocha2025}, which allows non-conservative mass transfer to occur when the star overflows its Roche lobe near periastron and tracks the resulting changes in orbital separation $a$ and eccentricity $e$ due to mass and angular momentum loss; (ii) an enhanced magnetic braking torque for cool stars, using prescriptions calibrated to recent models of stellar wind braking \citep{Garraffo2018}, in place of the standard over-simplified magnetic braking law \cite{Gossage2023}; and (iii) a state-of-the-art pulsar spin evolution module that follows the NS’s spin-up due to accretion torques and spin-down due to magnetic dipole radiation \citep{Chattopadhyay2020,Chattopadhyay2021}. We analyse the result of each binary under two limiting assumptions for the mass-transfer at RLOF: the standard, fully circularised orbit versus an eccentric orbital mass transfer case. By comparing these two scenarios for each system, we aim to bracket the extremes of how the Gaia
binaries might evolve: from no circularization (fully eccentric mass transfer) to complete circularization
at RLOF. In all cases, the mass-transfer rate is limited by the Eddington accretion rate onto the NS (we
also explored super-Eddington accretion in test runs, but found it did not qualitatively change the
outcomes, so we focus on the Eddington-limited results). We evolve each binary until the donor star has
lost its envelope and becomes a compact remnant (WD), and the NS’s accretion and
spin-up episode has ceased.
\begin{figure}  
  \centering
  \includegraphics[width=\linewidth]{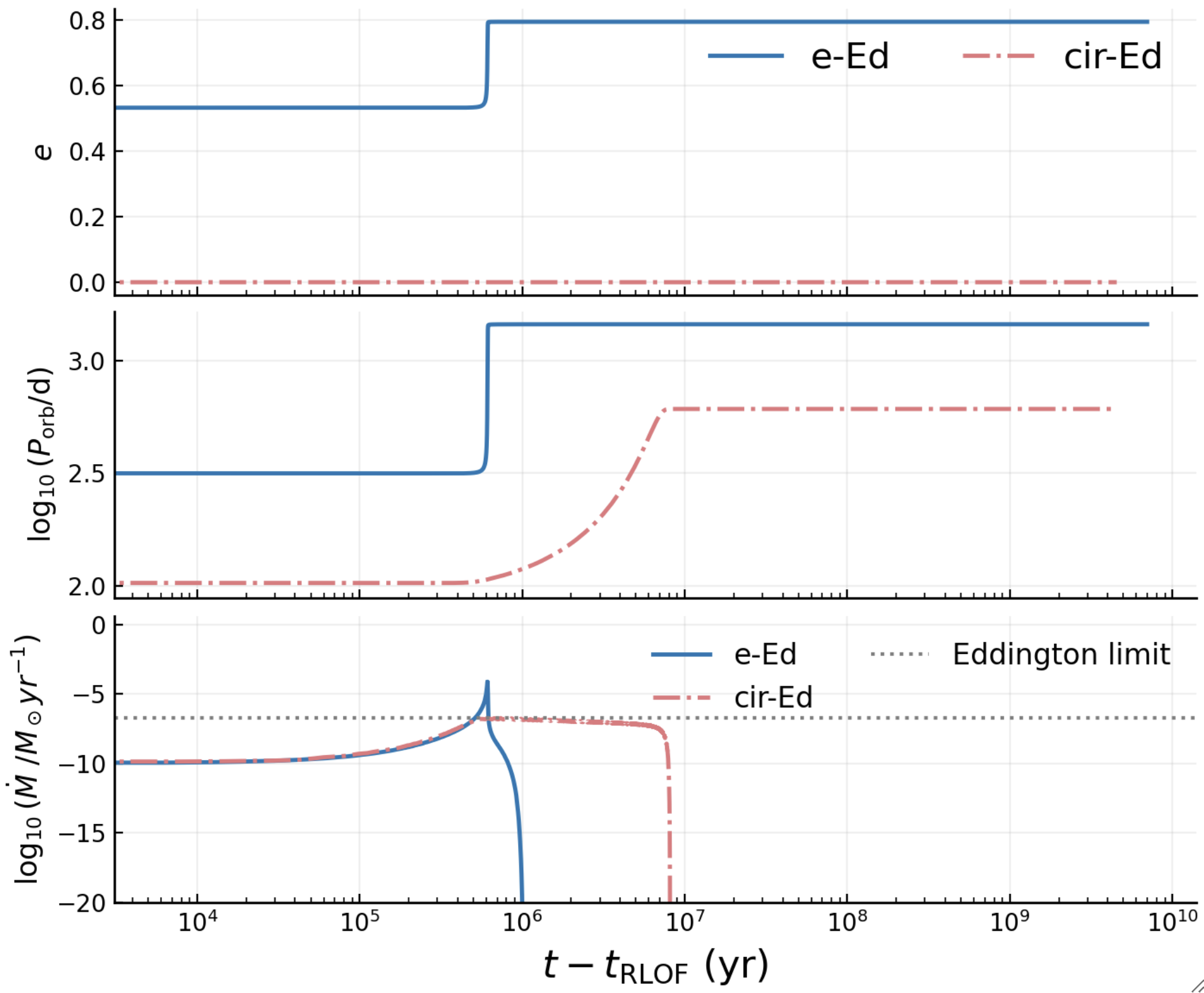}
  \caption{Time evolution of key binary and stellar properties for Gaia NS-MS \texttt{J1553-6846} with the NS mass at 1.323\,$M_\odot$, companion of 1.04\,$M_\odot$, $P_\mathrm{orb}=310.17\,d$, $e=0.5314$ at solar metallicity comparing the evolution for Eddington-limted circular (\texttt{cir–Ed}, pink ) and eccentric (\texttt{e-Ed}, blue) mass transfers. Panels from top to bottom show: (a) orbital eccentricity $e$; (b) $log_{10}(P_{\rm orb}/{\rm d})$; (c) mass-transfer rate log$_{10}(\dot M/M_\odot\,\mathrm{yr}^{-1})$.  All quantities are plotted versus time since RLOF, $t - t_{\rm RLOF} (yr)$.}
  \label{fig:plot1}
\end{figure}

\section{Results and Implications}
\subsection{Past with \texttt{POSYDON}}
Even under optimistic assumptions, we found that isolated
binary evolution produces Gaia-like NS–MS systems in at most about 1.5\% of
simulations. However, this negligible formation rate should be taken with a grain of salt. This is because while the relative rates of Gaia-like NS-MS systems are rare compared to the general NS-MS population that become NS-WDs, the normalized rates, if a continuous star formation rate is taken is consistent with the detections. Intriguingly, our synthetic populations also indicate that the majority of
``canonical” NS–WD binaries (the tight MSP–WD systems commonly observed) arise from a common envelope (CE) phase shortly after the NS’s birth. In other words, unstable mass transfer
is a key ingredient in the standard formation of MSP–WD binaries – a process the Gaia systems appear
to be avoiding in their future. While the time evolution from zero-age-main-sequence to canonical NS-WD formation appears to occur at a shorter timescale, owing to the fact that after NS forms in those cases, the companion evolving star appears to fall in the mass range of 2-8\,$M_\odot$, much more massive than the Gaia NS's sun-like companions. We defer further discussion of these formation aspects to future work, and proceed
with the assumption that regardless of their origin, the Gaia binaries will evolve in isolation going
forward.

\subsection{Evolutionary fate with \texttt{MESA}}

All 21 Gaia NS–MS binaries ultimately evolve into NS–WD binaries in our simulations, but the mass-transfer mode leads to two distinct evolutionary outcomes. Figure~\ref{fig:plot1} illustrates a representative example of the time evolution under each scenario, and the key differences are summarized below:

\noindent\textbf{Orbital evolution:} Under eccentric mass transfer, the orbit expands significantly during mass exchange—typically the final orbital period is about $3$–$4$ times the initial $P_\mathrm{orb}$ (for most systems). The eccentricity $e$ often remains high and in some models even increases slightly. By contrast, in the circular RLOF scenario, the orbit initially shrinks when mass transfer begins (as the orbit circularizes and mass-ratio effects drive tightening) but then widens gradually during the prolonged mass-transfer phase. Final orbital periods in the circular cases are usually only about $1.5$ times the initial period (and in a few cases the orbit ends slightly smaller than initially). All circular-case outcomes, of course, end with $e\simeq0$. The eccentric-transfer cases terminate as NS–WD binaries of substantial eccentricity (often $e\sim0.3$–$0.8$, roughly $1.5\times$ the initial $e$).
In both scenarios, we find that the loss of orbital angular momentum is governed almost entirely by mass loss (and the associated tidal coupling)
during and around the mass-transfer phase. Magnetic braking of the orbit and gravitational wave radiation are essentially negligible contributors in these wide, long-period systems. In the eccentric models, each periastron mass-loss burst removes a chunk of angular momentum, driving the orbital expansion. In the circular models, the continuous mass outflow carries away angular momentum. Our calculations confirm that tides plus mass loss account for the orbital evolution, whereas classical braking torques (strong in short-period compact binaries) do not
play a significant role.

\noindent\textbf{Mass-transfer duration and rate:} In the eccentric channel, mass transfer occurs in brief bursts at periastron. We find high peak mass-transfer rates of order $\dot{M}\sim10^{-5}$–$10^{-4}\,M_\odot\,\mathrm{yr}^{-1}$ at burst peak, but these intense episodes are short-lived, with the effective mass-transfer phase lasting $\lesssim10^6$\,yr before the donor detaches from its Roche lobe. In contrast, the circular RLOF cases exhibit a relatively steady mass-transfer rate around $\dot{M}\sim10^{-6}\,M_\odot\,\mathrm{yr}^{-1}$, persisting for a much longer period—up to $\sim10^7$\,yr—as the donor star is gradually stripped. Thus, eccentric mass transfer is intense but short, whereas circular mass transfer is gentler but sustained.
\begin{figure}
  \centering
  \includegraphics[width=0.8\linewidth]{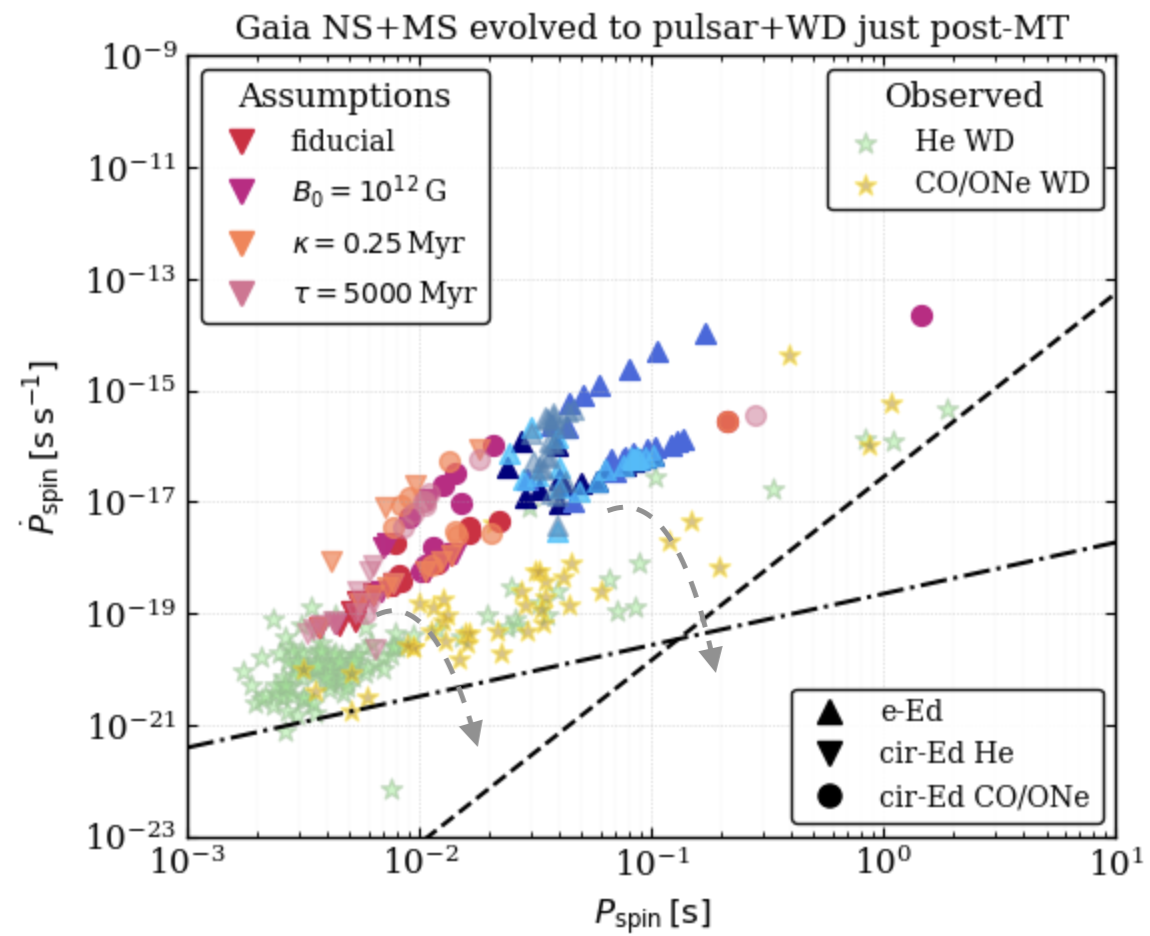}
  \caption{Spin‐period $P_\mathrm{spin}$ versus spin‐down rate $\dot{P}_\mathrm{spin}$ right at the end of stable mass transfer (MT)  for four representative pulsar models around `fiducial" (initial $P_\mathrm{spin}=50\,$ms, initial surface magnetic field $B_0=10^{11}\,$G, magnetic field decay timescale $\tau=$1000\,Myr, magnetic field decay mass scale $\kappa=0.025\,$Myr), and the following three with $B_0=10^{12}\,$G, $\kappa=0.025\,$Myr, $\tau=$5000\,Myr and all other parameters remain identical to fiducial (in different colour shades) chosen to maximise differences. Each model has 21 Gaia systems, \texttt{e-Ed} (blue, upright triangle, only has He WDs) and \texttt{cir-Ed} (pink, inverted triangle for He WD and circle for CO WD). The radio pulsar observations are depicted as green star (He WD) and gold star (CO/ONe WD). The grey arrows show the eventual time evolution of the pulsars since mass transfer that takes them towards the radio-quiet `death valley' bordered by the two pulsar `Death Lines'.}
  \label{fig:plot2}
\end{figure}

\noindent\textbf{White Dwarf Remnants:} In all simulations, the companion star ultimately loses its
hydrogen-rich envelope and becomes a WD. However, the core composition of the WD
differs by scenario. In every eccentric run, the donor’s envelope is shed relatively early (on the first ascent giant branch), leaving behind a low-mass helium (He) WD (He core that never grew beyond $\approx0.3\,M_\odot$). In the circular RLOF runs, some donor stars have time to evolve further and ignite helium before the envelope is fully removed. We find that 9 out of the 21 systems in the circular scenario develop carbon–oxygen (CO) cores, ending as higher-mass
CO WDs (~0.35–0.55 $M_\odot$) . The threshold for forming a CO WD in our models is a core mass $\gtrsim0.45\,M_\odot$ at the end of mass transfer , which typically occurs in systems with initially longer periods or more massive companions (allowing the donor to reach the He-burning stage during the extended mass-transfer). Notably, those circular-track
donors that form CO WDs sometimes undergo late helium shell flashes and thermal pulses,
creating loops in the Hertzsprung–Russell diagram (potentially observable as transient flashes by the Rubin observatory \citealt{LSST2019}). Such thermal pulses do not occur for the progenitors of He WDs.

\subsection{Connection  to Pulsar White Dwarfs}


A primary motivation of this study is to determine whether the Gaia-discovered NS–MS binaries are progenitors of the Galactic MSP–WD systems observed to date. Galactic radio surveys have detected of nearly a couple of hundred MSP–WD binaries in the field (excluding globular clusters). Almost all of these systems have circular or very low-eccentricity orbits, consistent with expectations for binaries that experienced long, stable phases of Roche-lobe overflow \citep{Bhattacharya1991,Kiziltan2010}. However, our \texttt{POSYDON} experiments already hint that many canonical NS–WDs may instead form through CE evolution rather than exclusively via stable mass transfer, consistent with growing theoretical \citep{MacLeod2015} and observational \citep{Yang2025} indications of significant pulsar recycling during the CE. While CE NS recycling has been studied by \citet{Chattopadhyay2020,Chattopadhyay2021} in the contexts of double-NS and NS–black-hole systems, a full population synthesis of NS–WD binaries—the largest observed class of binary pulsars—has yet to be carried out. Our initial \texttt{POSYDON} results provide a first indication of the importance of unstable mass transfer in producing MSPs. However, in our forward evolution of the 21 Gaia NS–MS systems, regardless of the mass-transfer prescription, the subsequent RLOF is stable. This leaves a degeneracy: prolonged, stable mass transfer under the circularized-at-RLOF assumption also produces MSPs in circular NS–WD binaries.

The pulsars in these systems are typically fast-spinning MSPs (periods of a few-30 milliseconds), and their WD companions span a range of masses—often $\sim0.2$–$0.3\,M_\odot$ He WDs, but also many in the $\sim0.4$–$0.5\,M_\odot$ range, indicating CO-core WDs in some cases. Observationally, there is a noted trend that pulsars with more massive WD companions tend to have slightly longer spin periods, whereas those with very low-mass He WDs are among the fastest-spinning MSPs. This empirical correlation has been discussed in previous population studies (e.g., \citealt{Kiel2008,Kiziltan2010}) and is generally interpreted as an imprint of how much accretion-driven spin-up the NS received: systems that produced heavy WDs likely experienced a shorter, late Case~B RLOF phase (spinning the pulsar up moderately), whereas those with lighter WDs underwent longer Case~A/B mass transfer and spun the pulsar up more fully.

Our circular RLOF models naturally reproduce these characteristics of the Galactic MSP–WD population. In the circular scenario, about half of the Gaia binaries produce low-mass He WDs and half produce higher-mass CO WDs; correspondingly, we find that NS spin periods are shorter for the He-WD cases and somewhat longer (tens of ms) for the CO-WD cases, matching the observed trend. The final orbital periods in the circular models (typically a few hundred days) and their tight orbital circularization are also fully consistent with known MSP binaries. In fact, some of the longer-period MSP–WD systems observed (with $P_{\rm orb}\sim200$–$800$\,d) could be direct analogues of the outcomes we predict under the circular channel. We conclude that stable, circular RLOF is the evolutionary route that links the Gaia NS–MS binaries to the majority of pulsar–WD systems in the field.

The eccentric mass-transfer channel does not appear to produce systems that match the typical observed pulsar–WD binaries. An eccentric mass-transfer outcome would be a wide binary (hundreds to thousands of days) with significant eccentricity ($e\sim0.3$–$0.7$), hosting a mildly recycled pulsar (spin periods of tens of ms to $0.1$–$1$\,s) and a low-mass He WD. Such a system would resemble an intermediate pulsar—slower than a MSP but spun up relative to a young pulsar—in a wide, eccentric orbit. These are observationally quite rare. In fact, no known Galactic field pulsar–WD binary has an eccentricity above $0.1$, except those in which the pulsar is young (not recycled) or the companion is not a WD (e.g., high-eccentricity pulsar B1913+16 has an NS companion \citealt{HulseTaylor1975}; eccentric PSR~J1903+0327 turned out to have a main-sequence companion \citealt{Champion2008}). The absence of eccentric MSP–He WD binaries could indicate that the eccentric mass transfer channel is a very minor contributor to the pulsar population—either because it seldom occurs (owing to the rarer occurrence of Gaia-like systems in the first place as predcited by \texttt{POSYDON}), or because the resulting pulsars are hard to detect. Selection effects likely play a role: a pulsar with spin period $P_\mathrm{spin}\sim0.1$–$1$\,s in a year-long eccentric orbit is difficult to identify in standard radio pulsar searches due to acceleration/sampling issues. Searches sensitive to long-period binaries have only recently improved (e.g., targeted Fast Fourier Transform algorithms and timing of candidates), and a few long-period MSPs have been found (e.g., the $P_{\rm orb}\sim669$\,d MSP 0407+1607 \citealt{LorimerFreire2005}). But a truly eccentric MSP binary would require careful acceleration or jerk searches \citep[e.g.,][]{Bagchi2013}. It is conceivable that some mildly recycled pulsars in eccentric orbits are lurking undetected in radio surveys due to these challenges \citep[see also][]{Chattopadhyay2021}. Additionally, the lifetime of a partially recycled pulsar may be shorter—if its magnetic field does not decay as much, its spin-down age could be relatively small, making it a dim radio source after a few $\times10^7$\,yr. All these factors suggest that even if the Gaia binaries do follow the eccentric mass-transfer route, the resulting pulsars would form a small and elusive subpopulation of Galactic pulsars.

Figure~\ref{fig:plot2} illustrates the recycled pulsars formed from the evolution of 21 Gaia binaries into NS–WD systems, for two stable mass-transfer cases—circularised at RLOF and eccentric—each evaluated with four different pulsar model variations, compared against the observed radio pulsar–WD systems.

It is worth noting that our eccentric mass-transfer modelling assumes an extreme limit: mass is transferred in sharp bursts at periastron, with effectively no mass flow during the rest of the orbit (a delta-function approximation). In reality, if a binary’s eccentricity is not too high (say $e\lesssim0.3$–$0.4$), tidal interactions may enforce partial circularization or phase-dependent mass transfer spread around periastron passage rather than instantaneous bursts. For those few Gaia systems with moderate eccentricities, our eccentric model might therefore overestimate how discontinuous the mass transfer is. A phase-dependent prescriptions by \citet{Hamers2019}—could yield somewhat different outcomes, potentially reducing eccentricity growth or allowing longer-lived transfer even in an eccentric orbit. Improving eccentric RLOF modelling is thus an important future step in refining predictions for the lower-eccentricity cases.

\section{Conclusions and Future Work(s)}

We have investigated the evolutionary fate of the newly discovered Gaia NS–MS binaries, emphasizing the impact of eccentric versus circular mass transfer on the formation of pulsar–WD binaries. Our findings can be summarized as follows.

\noindent\textbf{Gaia binaries as a rarer subpopulation:} Population synthesis with \texttt{POSYDON} indicates that Gaia-like NS–MS systems arise infrequently in isolated evolution—at most $\leq1.5\%$ under optimistic assumptions—and their normalized numbers are nonetheless consistent with current detections under a continuous star-formation history. Crucially, our synthetic populations suggest that the majority of canonical MSP–WD binaries are produced by unstable mass transfer leading to a CE episode that both circularizes the orbit and recycles the NS: brief but efficient CE–mediated accretion spins up the NS, setting the observed MSP periods and helping shape the MSP mass and luminosity distributions. In this sense, CE recycling emerges as a primary pathway to the Galactic MSP–WD population. Whereas, the Gaia NS–MS binaries appear to avoid a CE phase and, in forward evolution, undergo prolonged, stable RLOF—still capable of producing circular MSP–WDs, but likely contributing only a minority of systems given their rarity and the selection biases against discovering long-period, eccentric progenitors ($P_\mathrm{orb}\sim$ hundreds of days). Their wide, often eccentric orbits nonetheless provide a valuable laboratory for testing mass-transfer physics outside the CE channel.

\noindent\textbf{Mass-transfer geometry matters:} The mode of mass transfer leaves distinct imprints on the final binary. Eccentric mass transfer tends to produce wide, eccentric NS–He WD binaries with only mildly recycled pulsars, whereas circularized RLOF yields tighter (albeit still wider than the initial orbit), circular NS–WD binaries with fully recycled pulsars (often with CO-core WDs). The latter outcome closely resembles the bulk of observed Galactic MSP–WD binaries, suggesting that most MSP–WDs originate from the sustained, circular channel. In both eccentric and circular cases, orbital changes are dominated by mass and angular-momentum loss tied to the mass-transfer process itself. Neither magnetic braking nor gravitational radiation plays a significant role at these long orbital periods. This highlights how wide Gaia-like systems evolve under different physics than compact binaries.

\noindent\textbf{Neutron star accretion and spin-up:} Because of the limited duration and the fact that mass transfer in eccentric runs is highly super-Eddington only for short intervals (with most of the transferred mass lost from the system), the NS accretes relatively little in the eccentric mass transfer case—typically only a few $\times10^{-2}\,M_\odot$ in total. Consequently, the NS’s spin period only decreases to $P_\mathrm{spin}\sim50$–100 ms (a partially “recycled” pulsar). In some extreme eccentric cases the final spin could be hundreds of milliseconds or even approaching 1\,s, i.e., only mildly recycled. On the other hand, in the circular RLOF channel the NS accretes approximately an order of magnitude more mass—about $10^{-1}\,M_\odot$—owing to the longer-lasting, Eddington-limited accretion. This is sufficient to spin the NS up to millisecond periods. In our models, final spin periods of $P_{\rm spin}\approx 3$–30 ms are achieved in the circular cases (making the NS a fully recycled MSP). We note that these spin outcomes depend somewhat on assumed neutron-star birth spin and magnetic-field decay parameters, but across a broad range of plausible pulsar parameters, the trend holds: only sustained accretion can produce an MSP, whereas bursty periastron accretion yields a slower pulsar.

This work has focused on forward evolution given the Gaia-observed initial conditions. An equally important open question concerns their origin. Preliminary population-synthesis experiments with \texttt{POSYDON} show that the ``standard'' CE channel cannot readily account for them. Possible alternative formation routes include dynamical exchanges, disrupted triples, or other non-standard channels. We are actively exploring these scenarios. A forthcoming detailed paper will present a comprehensive synthesis of both formation and evolution, showing how unstable mass transfer produces the majority of MSP–WD binaries, while the Gaia sample represents outliers from a rarer channel.

In conclusion, Gaia’s NS–MS binaries serve as a natural experiment probing how mass transfer operates in eccentric binaries and how such systems may connect young neutron-star binaries to recycled pulsar–WD systems. Incorporating eccentric mass-transfer physics and pulsar spin evolution into binary models already yields testable predictions. With future Gaia data releases and upcoming pulsar surveys (e.g.\ SKA, MeerKAT), we will be able to confirm whether eccentric NS–WD systems populate the Galaxy in measurable numbers, or whether efficient tidal coupling typically drives them onto the circular recycling track. Together, these efforts will help build a unified evolutionary picture for NS binaries in our Galaxy, from formation to final MSP–WD outcomes.

\end{document}